# Determination of Redshifts for Selected IVS Sources. I.

**Kirill Maslennikov[1], Alexandra Boldycheva[1], Zinovy Malkin[1], and Oleg Titov[2]**

[1]Pulkovo Observatory, St. Petersburg, Russia
[2]Geoscience Australia, Canberra, Australia
contact e-mail: km@gao.spb.ru

**Abstract.** From observations with the 6-m BTA telescope at SAO RAS, we have determined spectroscopic redshifts of seven optical objects whose coordinates coincide with those of radio sources from the list of IVS (International VLBI Service for Geodesy and Astrometry). When compared to radio data, the obtained spectra and redshifts provide evidence for reliable identification of four observed objects; the other three require further study. The distances to the sources derived from our measurements will make it possible to refine the current estimates for parameters of cosmological models based on proper motions of these objects, which are determined from geodetic VLBI observations.

## 1. Introduction

In 1966, Kristian and Sachs [1] showed that within the framework of general relativity, anisotropic expansion of the Universe may result in systematic effects in proper motions of distant objects. These effects were described by quadratic vector spherical functions and were presented by expansion in the degrees of the distance to the objects r, the first term of which was formed by the group of summands multiplied by $r^0$, the second term by $r^1$, and so on. The authors restricted their consideration with the first two groups of the summands and suggested three possible explanations for these effects: anisotropic expansion of the Universe, rotation of the Universe, and primordial gravitation waves of poloidal and toroidal nature.

When the paper [1] appeared, it was impossible to detect the considered effects, since it was impossible to determine the proper motions with a sufficient accuracy (several microarcseconds (μas) per year). Presently, however, this accuracy is reached in VLBI measurements of proper motions of extragalactic radio sources. In these motions, preliminary estimates have revealed statistically meaningful harmonics described by a quadratic spherical function; they increase with the redshift [2, 3]. Systematic effects in this case increase from 10±3 μas/yr for sources within the redshifts interval 0 to 0.7 (with the average z = 0.44) to 25±8 μas/yr for sources with the redshifts 1.5 to 3 (the average z = 2.23). For sources with z > 1.7, a sharp increase in amplitudes of the proper motions was detected (58±10 μas/yr), which may be due to the lack of observations for sources with high redshifts.

Geodetic VLBI observations for radio sources have been carried out with radio telescopes located in different continents since the 1960-ies, however, at present, essentially in all astrometric works, only two-frequency observations are used, which have been carried out since 1979 and are stored in the IVS (International VLBI Service for Geodesy and Astrometry) database[1]. The total number of the observed radio sources is approaching 5000, but only around 1000 out of them were observed for a sufficiently long time (exceeding 10 years). For the most observed radio sources, with several hundred thousands observations within the time interval exceeding 30 years, the formal estimate for the coordinate accuracy is at the level of 10 μas, the real accuracy being about an order of magnitude higher [4].

---

[1] http://ivscc.gsfc.nasa.gov

Currently, apparent motions of radio sources, which, as a rule, are active galactic nuclei, are analyzed in the studies of instabilities in astrometric positions of reference objects. In addition to the abovementioned quadratic vector spherical harmonics, vector spherical harmonics of the first degree were also detected (a dipole and rotational components) with the amplitude of 10-20 µas/yr with the rms error of 1-2 µas/yr. The dipole component is apparently due to the accelerated motion of the barycenter of the Solar system around the center of the Galaxy [5-10], while the rotational component is probably due to the inaccurately determined constant of precession.

Up to now, is has been assumed that all apparent variations of the radio source coordinates are exceptionally due to the structure oscillations. However, the latest results indicate that the estimates of the proper motions also display systematic components. Indeed, when a source displays complex asymmetrical and evolving structure, its apparent (measured) coordinates are reduced to the position of the centroid of brightness of its radio image, while the structure variations caused by active processes in the core of the source result in variations in the apparent coordinates with the time. In particular, frequently observed relativistic motions of matter in AGN jets result in shifts in the centroids of the source radio brightness, and hence in the fictitious proper motions. Such proper motions reach several hundreds µas per year, are, as a rule, stochastic, and in a number of cases may alter their directions into the opposite within less than a year [11, 12]. This results in an increase in the random error in the estimates of vector spherical harmonics, as well as in the appearance of systematic errors. This effect may be softened by an increase in the number of the processed radio sources, which should be provided with sufficient observational history and, of course, with the redshifts. However, for most sources observed in geodetic VLBI surveys the redshifts are not determined [13, 14]. This deficit is more inherent in the southern celestial hemisphere, for the declinations $\delta < -40º$; however, for numerous sources in the northern hemisphere, the redshifts are still to be measured as well. Our study is particularly aimed at the increase in the number of astrometric radio sources with known distances (redshifts). It was suggested that, as a result, two interconnected problems related to the refinement in the available estimates for cosmological parameters derived from the analysis of proper motions may be solved:

     1. An increase in the number of radio sources with known proper motions and distances.
     2. An increase in the total number of geodetic radio sources with known distances, for their use in planned observational programs.

     The second aspect in our program of observations of distant radio sources for determination of cosmological parameters has been supported by the Program Committee of the International VLBI Service for geodesy and astrometry (OPC IVS). In October, 2007, OPC IVS approved the first three observational sessions of the program at the global IVS network, carried out in 2008. The results obtained in these sessions will be used to approve the continuation of the program. Thereby, our optical observations of radio sources are vital for the formation of the program of VLBI observations and the subsequent interpretation of its results.

## 2. The list of the objects

The determination of redshifts presents a difficult problem, which in our case includes optical identification of the sources. As most of the objects are faint, they are measurable only with large telescopes, which makes it hardly possible to provide redshift measurements for large samples of IVS objects. Therefore, it is important to separate priority sources, for which the determination of z is of particular interest. These are sources with the richest history of observations, since they are given the largest weights in the analysis of velocity fields of radio sources derived from astrometric VLBI observations. A systemized list of the priority sources was for the first time composed in 2007, and is being updated as VLBI observations are accumulated and new sources with the known redshifts appear. The latest version of the list is presented in [14].

When the observational time had been allocated at the BTA 6-m SAO RAS telescope in August, 2008, we selected from this list a sub-sample of objects that were the most convenient for observations at this time at the given latitude (Table 1).

Table 1. The list of the observed sources.

| IERS designation | RA hour, min, sec | DE deg, min, sec |
|---|---|---|
| 1751+288 | 17 53 42.4736 | +28 48 04.938 |
| 1923+210 | 19 25 59.6053 | +21 06 26.162 |
| 2013+163 | 20 16 13.8600 | +16 32 34.113 |
| 2023+503 | 20 25 24.9725 | +50 28 39.536 |
| 2030+547 | 20 31 47.9585 | +54 55 03.139 |
| 2152+226 | 21 55 06.4585 | +22 50 22.281 |
| 2302+232 | 23 04 36.4364 | +23 31 07.610 |

## 3. Observations: technique and results

The first series of our observations of optical spectra of IVS objects was carried out on Aug 24 and 28, 2008, with the BTA 6-m telescope at SAO RAS with the SCORPIO multimode spectrograph [15] in the long-slit mode. The width of the slit was 1"; a CCD matrix EEV-CCD42-40 (the size of the chip 2048x2048 elements, reading noise 1.8 electrons) was used as the detector. The spectra were obtained with the grism VPHG-550G in the wavelengths interval 3100-7300 Å with the instrumental resolution 10 Å for the inverse dispersion 2.1 Å/pix. The observational data were processed according to the standard technique with the use of MIDAS program package developed at ESO[2]. Along with the object spectra, reference spectra from a He-Ne-Ar lamp and the twilight sky were detected.

Figure 1 presents the obtained optical spectra of the objects, on which identified lines are indicated, while Table 2 contains the interpretation of these spectra and the measured redshifts. The columns of Table 2 present: 1 – the name of the object, 2 – identified spectral lines, 3 – wavelengths of the lines in the reference frame of the object, along with the observed wavelengths, 4 – the redshift, 5 – the object type specified, 6 – the date of the observations, 7 – the exposure time in minutes. Lines marked with the asterisk are observed in absorption, the others are emission lines. Below, we present our interpretation of the optical spectra and our classification of the objects.

---

[2] http://www.eso.org/sci/data-processing/software/esomidas/

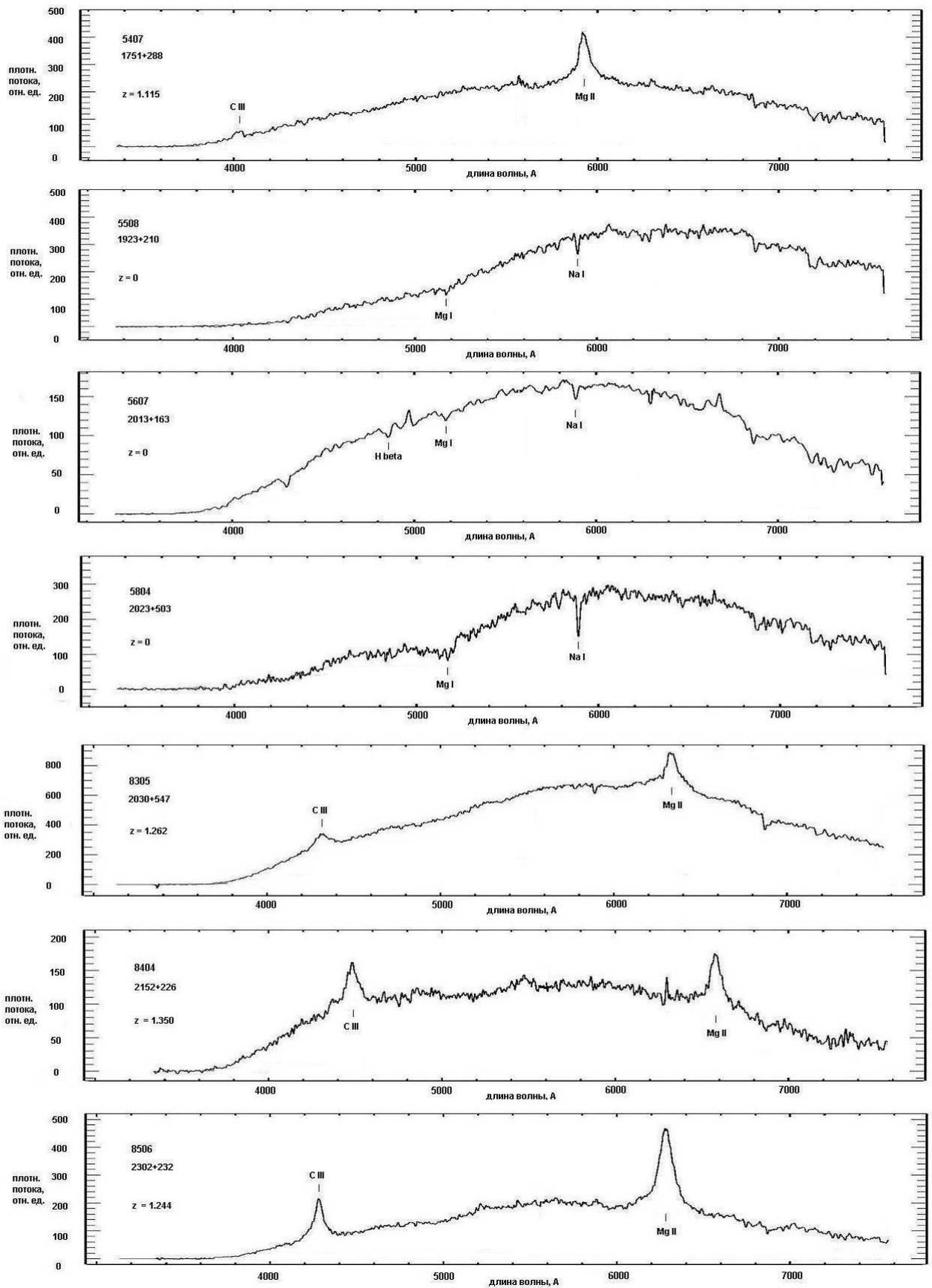

Рис. 1. Оптические спектры источников, сверху вниз: 1751+288, 1923+210, 2013+163, 2023+503, 2030+547, 2152+226, 2302+232.

## 3.1. IVS 1751+288

The spectrum displays bright C III 1909 and Mg II 2798 emission lines. The redshift is z = 1.115. The object is classified as a quasar.

## 3.2. IVS 1923+210

The spectrum of the object contains MgI 5170 and NaI 5893 absorption lines, which do not display any redshift. The object classification appears to be ambiguous: the VLBI data yield a pattern typical for a quasar with a clearly pronounced jet (Fig.2), which does not correspond to the obtained stellar spectrum. The NED database[3] notes that this is "an extended source at 327 MHz, possibly of the galactic nature", however, at the same time, it also identifies the source with the optical object that we observed, the single optical object seen in the vicinity of the radio source. The Westerbork Synthesis Radio Telescope 327 MHz Survey of the Galactic Plane [16] considers this object as intragalactic. Optically, a single object with the given coordinates is observed. A possible explanation for this discrepancy may be that the radio source is optically very faint (exceeding $22^m$), and accidentally coincides on the celestial sphere with a brighter star.

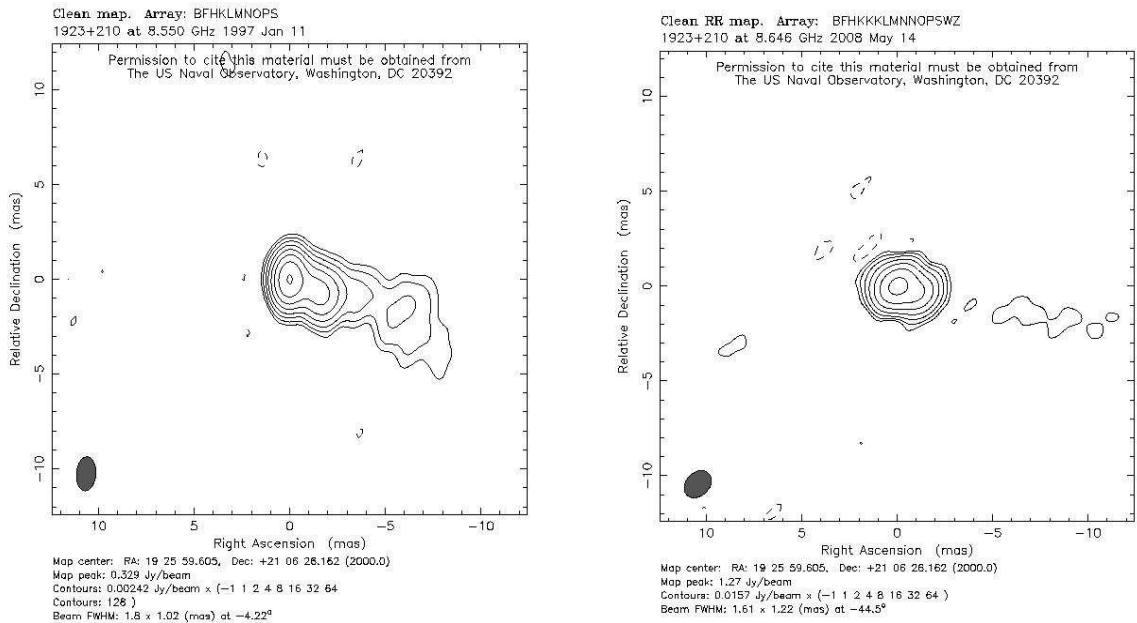

Fig. 2. Maps of the radio source 1923+210 from RRFID database[4] (courtesy of USNO).

## 3.3. IVS 2013+163

In the spectrum, Hβ 4861, MgI 5170, and NaI 5893 absorption lines are identified, without a redshift. Similar to the previous case, at radio wavelengths this source looks like a typical AGN with a clearly pronounced jet (Fig.3). The spectrum, however, corresponds to a star rather than to a galaxy or an AGN.

---

[3] http://nedwww.ipac.caltech.edu/
[4] http://rorf.usno.navy.mil/RRFID/

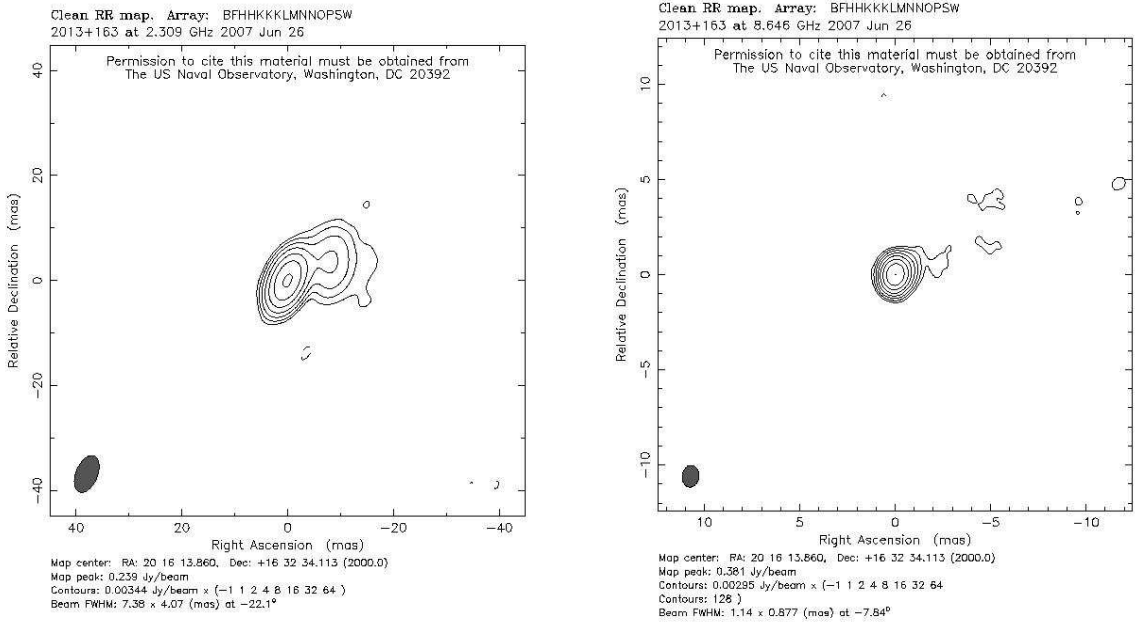

Fig.3. Maps of the radio source 2013+163 from RRFID database (courtesy of USNO).

### 3.4. IVS 2023+503

The spectrum displays unredshifted MgI 5170 and NaI 5893 absorption lines. In contrast to the previous two cases, at radio wavelengths this source does not present a typical AGN pattern, i.e., no pronoumced jet is seen (Fig.4), which may be due to insufficient sensitivity of the radio observations (Yu. Kovalev, private communication). We classified the obtained spectrum as stellar. No other optical objects brighter than~$20^m$ is detected in the vicinity of the radio source.

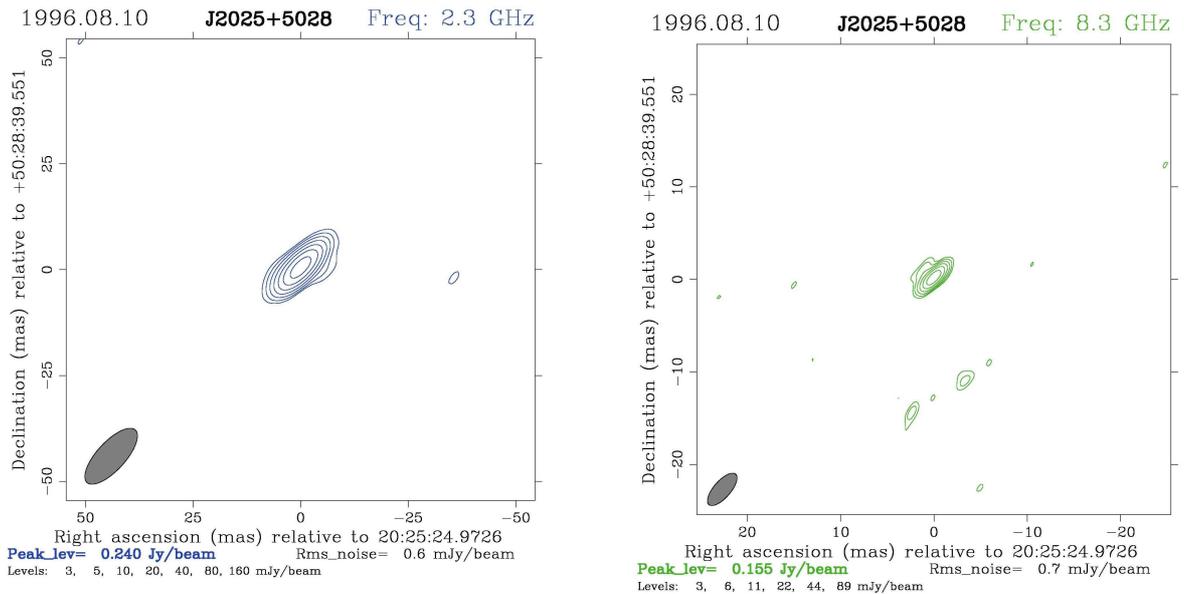

Fig.4. Maps of the radio source 2023+503 from L.Petrov's database[5] (courtesy of the author of the maps, Yu. Kovalev).

---
[5] http://astrogeo.org/vlbi_images/

### 3.5. IVS 2030+547

The spectrum of a rather bright 2030+547 source displays broad CIII 1909 and MgII 2798 emission lines, according to which its redshift is z=1.262. At this redshift, the presence of only these two lines is characteristic for quasars.

### 3.6. IVS 2152+226

The spectrum of the source contains CIII 1909 and MgII 2798 emission lines; the redshift is z=1.350. The object is classified as a quasar.

### 3.7. IVS 2302+232

The radio source is identified as a quasar, with 2 bright CIII 1909 and MgII 2798 emission lines in the optical spectrum. The measured redshift for the source is z=1.244.

## 4. Conclusion

We have obtained optical spectra for seven objects that presumably correspond to radio sources from the IVS program. For four sources, their spectra display bright emission lines typical for quasars, and noticeable redshifts. Therefore we conclude that these optical objects are reliably identified with the radio sources. The estimated accuracy of our determinations for z is 0.001. The other three observed objects display the spectra typical for stars, with the radial velocities close to zero with the same accuracy. This is in disagreement with their radio mapping, according to which these sources display the structure typical for AGN. This disagreement may be explained by overlapping on the celestial sphere a faint extragalactic object and a galactic star.

It should be noted, however, that the inconsistency between the typical stellar spectra and the radio structure inherent in extragalactic objects was manifested for three out of seven program objects, which is unlikely to be a mere coincidence. The suggestion that in all these cases an optically faint object and a star are randomly projected onto the same point of the sky (taking into account that no other optical objects are seen within 10-15") does not seem realistic, either. Note also that in the course of the selection of the objects for our spectral observations, which we carried out using the list [14] and DSS, we repeatedly experienced the total absence of any visible DSS objects at the coordinates for the IVS radio sources (this may be exemplified with the sources IVS 1932+204, IVS 1922+155, IVS 1955+335, and also, possibly, IVS 2000+148 and IVS 1951+355 - these five "empty fields" were found in a random sample of 30 IVS objects). This problem seems to be worth studying.

Here, we made the first step to the solution of the general problem of the determination of the distances to all, or at least to all most frequently observed radio sources of the IVS program. However, the list of the latter is far from exhaustion, therefore this study will be continued provided the observation time is available. To make the observations less time-consuming, relatively less cumbersome, though less accurate, massive photometrical observations for the determination of the redshifts might be carried out with smaller telescopes. The accuracy for z in this case may reach z 0.03-0.1. Later on, with larger telescopes, the redshifts may be refined for the most important objects.

## 5. Acknowledgements

The authors thank Dr. S. N. Dodonov (SAO RAS), who supervised the BTA observations and made useful comments in the classification of the obtained spectra.

In this work, we used the data from the RRFID database of images of radio sources of the USA Naval Observatory (USNO) and the NED database of the Jet Propulsion Laboratory (JPL), USA, and also the maps of radio sources obtained by Yu. Kovalev within VCS program [17-22].